\documentclass[12pt, preprint]{aastex}

\newcommand{\Mdot}{M$_{\sun}$ yr$^{-1}$}

\shortauthors{Brittain, S. et al.}
\shorttitle{A Brief Outflow from V1647 Ori}

\begin{document}

\title{Post-Outburst Observations of V1647 Ori: Detection of a Brief  Warm, Molecular Outflow}

\author{Sean Brittain}
\affil{Department of Physics and Astronomy, Clemson University, Clemson, SC 29634-0978}
\email{sbritt@clemson.edu}

\author{Terrence W. Rettig}
\affil{Center for Astrophysics, University of Notre Dame, Notre Dame, IN 46556}

\author{Theodore Simon}
\affil{Institute for Astronomy, University of Hawaii, 2680 Woodlawn Dr, Honolulu, HI 96822}

\author{Dinshaw S. Balsara, David Tilley}
\affil{Center for Astrophysics, University of Notre Dame, Notre Dame, IN 46556}

\author{Erika Gibb}
\affil{University of Missouri at St. Louis, 8001 Natural Bridge Road, St. Louis, MO, 63121}

\author{Kenneth H. Hinkle}
\affil{National Optical Astronomy Observatory P.O. Box 26732, Tucson, AZ 85726-6732}

\begin{abstract}
We present new observations of the fundamental ro-vibrational CO spectrum of V1647 Ori, the young star whose recent outburst illuminated McNeil's Nebula. Previous spectra, acquired during outburst in 2004 February and July, had shown the CO emission lines to be broad and centrally peaked---similar to the CO spectrum of a typical classical T Tauri star.   In this paper, we present CO spectra acquired shortly after the luminosity of the source returned to its pre-outburst level (2006 February) and roughly one year later (2006 December and 2007 February). The spectrum taken in 2006 February revealed blue-shifted CO absorption lines superimposed on the previously observed CO emission lines. The projected velocity, column density, and temperature of this outflowing gas was 30 km s$^{-1}$, $3^{+2}_{-1}\times10^{18}$ cm$^{-2}$, and 700$^{+300}_{-100}$ K, respectively. The absorption lines were not observed in the 2006 December and 2007 February data, and so their strengths must have decreased in the interim by a factor of 9 or more.  We discuss three mechanisms that could give rise to this unusual outflow.  
\end{abstract}


\clearpage

\section{Introduction}
McNeil's Nebula was recently illuminated by the outburst of V1647 Ori (McNeil 2004), a young star that is embedded in the Lynds 1630 dark cloud and coincides with the 850$\micron$ continuum source OriBsmm55 (Mitchell et al. 2001). V1647 Ori has a flat SED in the mid-infrared, and is thus a Class I young stellar object (YSO; Andrews et al. 2004). V1647 Ori underwent a similar outburst as recently as 1966 (Aspin et al. 2006), indicating that V1647 Ori is also an EXor. Such pre-main sequence stars undergo eruptive events that dramatically increase their luminosity for periods of months to years (Hartmann 1998). The outbursts are thought to be triggered by a rapid increase in the stellar accretion rate (Hartmann \& Kenyon 1996).  The eruption in November 2003 of V1647 Ori lasted two years.  During the outburst, the star brightened by a factor of 50 in  X-rays (Kastner et al. 2006), a factor of 250 in the red (6 mag in the $R_C$ band; Fedele et al. 2007a; Brice\~{n}o et al. 2004), a factor of 15 in the near-IR($\sim$3 mag in the $J$, $H$, and $K$ bands; Reipurth \& Aspin 2004), and a factor of $\sim$15 at wavelengths from $3.6\micron$ to $70\micron$ (Muzerolle et al. 2005). From the overall brightening of the source, Muzerolle et al. (2005) concluded that the bolometric luminosity increased by a factor of 15 to 44$L_\sun$ (see also Andrews et al. 2004) and that the stellar accretion rate increased from $\sim10^{-7} M_{\sun}$ yr$^{-1}$ to $\sim10^{-5} M_{\sun}$ yr$^{-1}$.  Similarly, Gibb et al. (2006) inferred a stellar accretion rate of $3-6\times10^{-6} M_{\sun}$ yr$^{-1}$ from the luminosity of the Br$\gamma$ emission one year later. This is somewhat larger than the typical accretion rate of a young low mass star ($10^{-8}-10^{-7} M_{\sun}$ yr$^{-1}$; Bouvier et al. 2007), yet lower than is expected for a star of the FUor type ($\sim10^{-4} M_{\sun}$ yr$^{-1}$; Hartman \& Kenyon 1996).

During the onset of the outburst of V1647 Ori, observations of atomic lines with P Cygni profiles provided evidence for a hot
($T\sim$10,000 K), high velocity ($v = -400$ km s$^{-1}$) wind (Brice\~{n}o et al. 2004; Reipurth et al. 2004; Vacca et al. 2004; Walter et al. 2004; Ojha et al. 2006; Fedele et al. 2007a) with a mass-loss rate of $\dot{M}_{\rm wind}$=4$\times$10$^{-8}$ \Mdot (Vacca et al. 2004). This mass loss rate is much lower than that of the typical FUor (Hartmann \& Kenyon 1996) and comparable to that of a classical T Tauri star (cTTS; Hartigan et al. 1995). The absorption component of the P Cygni profile of several lines (e.g. Pa$\beta$) disappeared within a few months following the peak of the outburst in early 2004 (Gibb et al. 2006). However, P Cygni structure in the H$\alpha$ profile indicated that a weaker wind continued throughout the outburst phase (Ojha et al. 2006; Fedele et al. 2007a).  

In contrast to the hydrogen and helium lines,  the fundamental near-infrared ro-vibrational emission lines of CO, observed 2004 February 27, were broad, centrally peaked, and compatible in their intensity with an excitation temperature of 2500 K (Rettig et al. 2005).  The width of the lines was shown to be consistent with Keplerian orbital motion of the gas within the inner disk surrounding the central star, similar to the broad emission line profiles that are observed around cTTSs and Herbig Ae/Be stars (HAeBes; Najita et al. 2003; Blake \& Boogert 2004).  A later observation, obtained on 2004 July 30, showed that the CO lines remained broad but the temperature of the gas decreased to 1700 K (Gibb et al. 2006).   Neither observation showed any indication of CO in an outflow, as a blue shifted absorption component was not detected.

We report followup observations of CO from V1647 Ori, which were acquired 2006 February, 2006 December, and 2007 February. By the time of our initial 2006 observation, V1647 Ori had returned to quiescence and the absorption component in the H$\alpha$ line profile had disappeared (Fedele et al. 2007a and references therein). Presumably the accretion rate had fallen by two orders of magnitude to its pre-outburst accretion rate as the star faded to its pre-outburst brightness (see Muzerolle et al. 2005).  As suggested by the work of Najita et al. (2003),  only a continued decrease in CO line intensity from the warm gas in the disk was therefore expected.  However, the first post-outburst observation revealed the striking metamorphosis of these lines from centrally peaked emission features to emission lines with blue-shifted absorption. Subsequently, by late 2006 and early 2007, the CO emission lines returned to their original centrally peaked structure, indicating that the production of the outflow diminished within one year of the end of the outburst.  In this paper we discuss  three scenarios that can give rise to such a phenomenon.

\section{Observations \& Data Reduction}
V1647 Ori was observed with NIRSPEC at the W. M. Keck Observatory on 2006 February 17 using one $M$-band setting (Table 1; Figs. 1 and 2). NIRSPEC is a high-resolution ($\lambda/\delta\lambda\sim25,000$), near-infrared (1--5 $\micron$), cross-dispersed spectrometer (McLean et al. 1998). The NIRSPEC data were reduced and calibrated using standard reduction techniques, which are described in DiSanti et al. (2001),  Brittain et al. (2003) and Brittain (2004). Additional observations of V1647 Ori were taken with the Phoenix spectrometer on Gemini South on 2006 December 01, 2007 February 01, and 2007 February 06, using one setting centered at 2011 cm$^{-1}$ (Table 1). Phoenix is a high-resolution ($\lambda/\delta\lambda\sim50,000$), near-infrared (1--5$\micron$) spectrometer covering 1550 km s$^{-1}$ per observation (Hinkle et al. 1998; Hinkle et al. 2000; Hinkle et al. 2003).  The same reduction algorithms used on the NIRSPEC data were applied to the Phoenix data. 

Spectra of the P30 and P31 CO lines from the four epochs spanning 2004--2007 are presented in Figure 1 in order to illustrate how the CO profiles have evolved over time. Gaps in the spectra represent frequencies with less than 50\% atmospheric transmittance.  The sample error bars presented for each spectrum represent the standard deviation of the continuum. Because the individual Phoenix spectra have lower S/N, we plot just the average of those datasets, weighting each observation by the standard deviation of the continuum. The equivalent widths of the emission and absorption lines measured from the new datasets are presented in Table 2.  Upper limits on the absorption lines in the Phoenix spectrum were calculated by adopting the width of the absorption lines observed in 2006 February, setting the depth of the line equal to the standard deviation of the continuum, and assuming a gaussian line profile. 

\section{Results}
The barycentric velocity of L1630 is +26 km s$^{-1}$ (Lada, Bally, \& Stark 1991; Gibb 1999; Mitchell et al. 2001). Since the velocity of young embedded stars is typically within a few km s$^{-1}$ of the surrounding cloud, we adopt this as the barycentric velocity of V1647 Ori. The emission lines, corrected for the motion of the earth, are centered on the barycentric velocity of the star under the assumption that it has the same velocity as L1630 (Fig. 1).

Rettig et al. (2005) noted that the unblended CO emission lines observed during the early phase of the outburst of V1647 Ori were symmetric and centrally peaked, hence they concluded that the CO emission originated in the inner disk, as in cTTSs and HAeBes.  The CO emission lines from these sources are generally centrally peaked, although a few of them exhibit double-peaked structure due to rotational broadening.  The broadening of the emission lines, the absence of blue-shifted absorption, and the excitation temperature  indicate that the emitting gas around cTTSs and HAeBes lies within a circumstellar disk (Najita et al. 2000).  Our 2006 February observation of V1647 Ori, however, reveals the surprising appearance of a blue-shifted absorption feature superimposed on the CO emission lines.  This absorption feature was blue-shifted from line center by 30  km s$^{-1}$ and had a width of 20 km s$^{-1}$. Roughly one year later, in 2006 December and 2007 February, the strength of the absorption decreased by at least a factor of 9 in such prominent lines as P30 (Table 2) and was no longer observed (Fig. 1). 

An excitation plot for the CO emission lines observed in 2006 February (after the star had returned to its preoutburst brightness) implies that the effective rotational temperature of the emitting gas was 1400$\pm$200 K. To estimate the temperature of the absorption lines, we fit the observations with a synthetic spectrum, which we calculate by assuming the lines have a gaussian profile, the $^{12}$CO/$^{13}$CO ratio is 60, and the gas is characterized by a single temperature. The free parameters in our model are the intrinsic line width, $b$, the CO column density, N(CO), and the rotational temperature of the gas, $T$.  We present the predicted equivalent widths of $^{12}$CO absorption lines in Table 2 and three models in Figure 2. The parameters for each model are as follows. Red spectrum: $T=1000$K, $b=2.5$ km s$^{-1}$, N(CO)$=2\times$10$^{18}$ cm$^{-2}$; green spectrum: $T=700$K, $b=2.0$ km s$^{-1}$, N(CO)=$3\times10^{18}$ cm$^{-2}$; and blue spectrum, $T=600$K, $b=1.8$ km s$^{-1}$, N(CO)=5$\times$10$^{18}$ cm$^{-2}$. The three sets of parameters do an equally good job of fitting the low-$J$ $^{12}$CO lines. If the gas is 1000K or hotter, the high-$J$ lines are too deep. If the gas is 600K or cooler, the low-$J\ ^{13}$CO lines from 2127--2151 cm$^{-1}$ (R9--R15) become too strong.  Our 1$\sigma$ upper limit on the observed equivalent widths of $^{13}$CO lines not blended with other features is 0.006 cm$^{-1}$.  For gas at 600K, 700K and 1000K,  the predicted strengths of these lines are $\sim$0.018 cm$^{-1}$, $\sim$0.011 cm$^{-1}$ and $\sim$0.006 cm$^{-1}$, respectively.  The model with $T=700$K (green) provides the best overall fit, so we conclude that $T=700^{+300}_{-100}$ K, N(CO)$=3^{+2}_{-1}\times10^{18}$ cm$^{-2}$, and $b=2.0^{+0.5}_{-0.2}$ km s$^{-1}$.

\section{Discussion}
During quiescence, V1647 Ori is a class I YSO (Andrews et al. 2004). However, its mass-loss rate during outburst was similar to a strongly accreting cTTS (Vacca et al. 2004). While outflows from cTTSs and HAeBes are common, fundamental ro-vibrational CO emission lines with a blue-shifted absorption component have not been observed around any of the more than 300 such sources observed to date (Najita et al. 2000, 2003; Blake \& Boogert 2004; Rettig et al. 2006; Brittain et al. 2007; J. Brown private communication). Further, Class I YSOs such as GSS 30 IRS 1, HL Tau and RNO 91 do not show ro-vibrational CO emission lines with blue-shifted absorption components (Pontoppidan et al. 2002, Brittain et al. 2005, and Rettig et al. 2006, respectively). In contrast to these systems, the FUor V1057 Cyg has shown blue-shifted overtone ro-vibrational CO absorption lines (Hartmann et al. 2004).  While the similarity of the unusual CO outflows is intriguing, there are important differences between FUors such as V1057 Cyg and EXors such as V1647 Ori. First, FUors have greater accretion rates ($\sim$10$^{-4}$ \Mdot) and more extreme winds than EXors (Hartmann \& Kenyon 1996). Secondly, CO is always and {\it only} detected in absorption toward FUors, and is thought to originate in the accretion disk (Calvet et al. 1993; Calvet et al. 1991). Despite the significant differences between V1057 Cyg and V1647 Ori, both stars have undergone major eruptions that have generated outflows in CO, suggesting that a different mass-loss mechanism may be at work in these systems than in other young stars. 

Given the unusual nature of the outflowing CO from V1647 Ori, it is of interest to consider the mechanisms that could give rise to this phenomenon. One possibility is that the absorbing CO condensed out of the hot outflowing wind that was observed during the outburst.  There were two outflow components noted in the H$\alpha$ absorption feature: a variable component at 400 km s$^{-1}$ and a steady component at 150 km s$^{-1}$ (Fedele et al. 2007a).  If the lower-velocity gas decelerated at a constant rate to $30$ km s$^{-1}$ (the velocity of the outflowing  CO), by the time of our second CO observation in 2004 July the wind would have expanded to 10 AU. There was no evidence of blue-shifted CO absorption on this date (Fig. 1). By 2006 February, when the absorption was observed, the moderate velocity wind would have expanded to nearly 40 AU. However, it seems unlikely that the wind could still retain a kinetic temperature of 700K at a distance of 40AU from the star. Furthermore, it is not clear why CO would condense out of this outflow to reveal warm, blue shifted absorption but not out of any of the other outflows with similar or even greater mass-loss rates. Thus we conclude that this scenario is unlikely.

A second possibility is that the CO absorption formed in a shell of material that was swept up by the atomic wind.  In this case the absorption did not appear until the column density of material was sufficient to produce measurable absorption, and the heating was the result of the interaction of the wind with the shell. When the mass loss decreased at the end of the outburst phase, this heating was eventually shut down.  If the CO/H$_2$ ratio in such a shell was similar to that of a dense molecular cloud, 1.5 $\times$10$^{-4}$, then the column density of gas was 2$\times$10$^{22}$ cm$^{-2}$. Adopting a normal interstellar extinction--to--gas ratio, i.e., $A_V=5.6\times 10^{22}  N_{\rm H}$ mag cm$^2$ atom$^{-1}$ (Bohlin et al. 1978), we find that the observed column density of CO corresponded to $\sim$20 mags of visible extinction. This is a lower limit, as it is possible that selective dissociation of gas in the nebula could drive down the relative abundance of CO.  However, the extinction measured on the line of sight toward V1647 Ori appears to have remained relatively unchanged over the entire course of the outburst at $\sim$11 mags (Vacca et al. 2004; Gibb et al. 2006), of which 6.5 mags is due to the nebula (Fedele et al. 2007b).  There is no evidence for an additional 10--20 mags of extinction toward V1647 Ori, and so we conclude that this scenario is also unlikely. 

A final scenario we consider is that the CO outflow was launched in response to the reorganization of the stellar magnetic field following the sharp drop in the accretion rate.  Pre-main sequence stars tend to have kilogauss magnetic fields which mediate stellar accretion and outflows (e.g. Johns-Krull et al. 2000).  This stellar magnetic field truncates the accretion disk where the ram pressure from accretion balances the magnetic pressure from the magnetosphere (Camenzind 1990).  Consequently, the truncation radius of the disk, $R_T$, is inversely and nonlinearly proportional to the accretion rate, $\dot{M}$, and given by $ R_T \propto \dot{M}^{-2/7}$ (e.g. Bouvier et al. 2007).  Thus the two order of magnitude change in the stellar accretion rate experienced by V1647 Ori would have resulted in the truncation radius being shifted by nearly a factor of four. 
 
Two years into the outburst, V1647 Ori rapidly faded by a factor of 40 in the R$_C$-band in just 180 days to return to its pre-outburst level (Fedele et al. 2007a).  The mid-infrared flux also returned to its pre-outburst level, indicating that the drop in the 
optical/near-infrared lightcurve was due to the intrinsic fading of the source (Fedele et al. 2007a). 
This sharp drop in the luminosity of the source indicates that the accretion rate fell rapidly as the star returned to quiescence. In response to the drop in ram-pressure from the accretion flow, the truncation radius was pushed back by the magnetosphere. Evidence from simulations suggests that disk-magnetosphere systems tend to form outflows when the system is undergoing the greatest amount of dynamical rearrangement (e.g., Fig. 7 in Balsara 2004). It is possible that the realignment of the magnetic field as it pushed out against the circumstellar disk resulted in a warm, shortlived outflow, an outflow that is not observed toward any other cTTSs or HAeBes.

While virtually all accreting low-mass stars drive outflows, the CO outflow from V1647 Ori is highly unusual. Indeed, the transformation of centrally peaked ro-vibrational CO emission lines to CO emission lines with blue-shifted absorption is unique.  We suggest that the mechanism responsible for producing this outflow is distinct from the one that drives the outflow from typical cTTSs. The coincidence between the rapid fading of V1647 Ori and the subsequent observation of the CO outflow hints at a connection. Better sampling of the fundamental ro-vibrational CO spectrum of EXors as they brighten and fade is crucial for determining whether this coincidence is significant. The rapid and dramatic change in the accretion rate that characterizes the EXor phenomenon provides an important opportunity to study the interplay between stellar accretion, the inner disk, and outflows. This insight is key to reaching a satisfactory theoretical understanding of these events.

\acknowledgements
The data presented herein were obtained (in part) at the W.M. Keck Observatory, which is operated as a scientific partnership among the California Institute of Technology, the University of California and NASA. The Observatory was made possible by the generous financial support of the W.M. Keck Foundation. 
Also based in part on observations obtained at the Gemini Observatory, which is operated by AURA, under a cooperative agreement with the NSF on behalf of the Gemini partnership. The Phoenix infrared spectrograph was developed and is operated by NOAO. 
 The Phoenix spectra were obtained as part of program GS-2006A-DD-1 and GS-2006B-DD-1.
S.D.B. performed this work (in part) under contract with JPL funded by NASA through the Michelson Fellowship Program.

\clearpage

\begin{deluxetable}{lcccc}
\tablenum{1}
\tablewidth{0pt}
\tablecaption{Log of Observations}

\tablehead{ \colhead{Date} & \colhead{Telescope/Instrument} &
\colhead{Spectral Grasp}  &  \colhead{Integration} & \colhead{S/N} \\
\colhead{}  & \colhead{} & \colhead{(cm$^{-1})$} & \colhead{(seconds)} & {}}
\startdata
2006 Feb 17 &  KECK 2/NIRSPEC & 2151--2119, & 480 & 29 \\
  &  & 2018--1987&  & 28\\
2006 Dec 01 & Gemini S./PHOENIX & 2016--2006 & 960 & 6  \\
2007 Feb 01 & Gemini S./PHOENIX & 2016--2006 & 4800 &  11 \\
2007 Feb 06 & Gemini S./PHOENIX & 2016--2006 & 3360 &  8 \\
\enddata
\end{deluxetable}

\clearpage

\begin{deluxetable}{lccccccc}
\tablenum{2}
\tablewidth{0pt}
\tablecaption{Equivalent Widths of Observed $^{12}$CO Lines and Model Predictions\tablenotemark{a}}

\tablehead{ \colhead{Transition} & \colhead{$\tilde{\nu}_{rest}$}  & \colhead{E.W.\tablenotemark{b}} &
\colhead{E.W.\tablenotemark{c}} & \colhead{E.W.\tablenotemark{c}} & \colhead{E.W.\tablenotemark{c}} &\colhead{E.W.\tablenotemark{b}}   \\
\colhead{} & \colhead{} & \colhead{absorption} & \colhead{$T=1000$K} & \colhead{$T=700$K} & \colhead{$T=600$K} & \colhead{emission} & \\
\colhead{} &  \colhead{(cm$^{-1}$)} & \colhead{(10$^{-2}$ cm$^{-1}$)} & \colhead{(10$^{-2}$ cm$^{-1)}$} & \colhead{(10$^{-2}$ cm$^{-1}$)} & \colhead{(10$^{-2}$ cm$^{-1}$)} & \colhead{(10$^{-2}$ cm$^{-1}$)}}
\startdata
\multicolumn{7}{c}{a) NIRSPEC (2006 February)} \\ \tableline
R1 & 	2150.86	&	4.41	$\pm$	0.60	&	4.52	&	4.69	&	4.81	&	(3.4) & \\
R0 & 	2147.08	&	3.49	$\pm$	0.60	&	3.30	&	3.98	&	4.29	&	(3.4) & \\ 
P1 & 	2139.43	&	3.40	$\pm$	1.00	&	3.27	&	3.96	&	4.26	&	(1.7) & \\
P2 & 	2135.55	&	5.17	$\pm$	0.80	&	4.47	&	4.64	&	4.77	&	(1.7) & \\
P3 & 	2131.63	&	4.33	$\pm$	0.60	&	5.05	&	4.97	&	5.02	&	(2.2) & \\
P4 & 	2127.68	&	5.03	$\pm$	0.60	&	5.39	&	5.18	&	5.17	&	-3.64$\pm$1.72 & \\
P5 & 	2123.70    & 	6.21$\pm$	0.60	&      	5.62	&	5.31 &	5.28 &	-3.59$\pm$1.72 & \\	
P6 & 	2119.68    & 	5.09$\pm$	1.00	&     	5.79 &	5.40 &	5.35 &	 -3.74$\pm$1.71 & \\	
P30 & 	2013.35	&	3.57	$\pm$	0.36	&	4.45	&	3.42	&	3.25	&	-3.80$\pm$0.39 & \\
P31 & 	2008.53	&	2.42	$\pm$	0.24	&	4.23	&	3.15	&	2.97	&	-4.05$\pm$0.35 & \\
P32 & 	2003.67	&	1.44	$\pm$	0.94	&	4.00	&	2.85	&	2.66	&	-3.99$\pm$0.34 & \\
P33 &	1998.78	&	2.86	$\pm$	0.64	&	3.75	&	2.53	&	2.32	&	-2.22$\pm$0.41 & \\ \tableline

\multicolumn{7}{c}{} \\  [-11pt]
\multicolumn{7}{c}{b) PHOENIX (2006 December and 2007 February) } \\   \tableline
P30 & 	2013.35 &	 	(0.4) 	& ... & ... & ...		& 	-2.35$\pm$0.31& \\
P31 &	2008.53 &	 	(0.8)	   & ... & ... & ...      		& 	-1.07$\pm$0.63 & \\ 
\tableline
\enddata
\tablenotetext{a}{Upper limit on measured equivalent width denoted by ().}
\tablenotetext{b}{Observed value.}
\tablenotetext{c}{Predicted value.}

\end{deluxetable}

\clearpage
\begin{figure} 
\epsscale{1.0}
\plotone{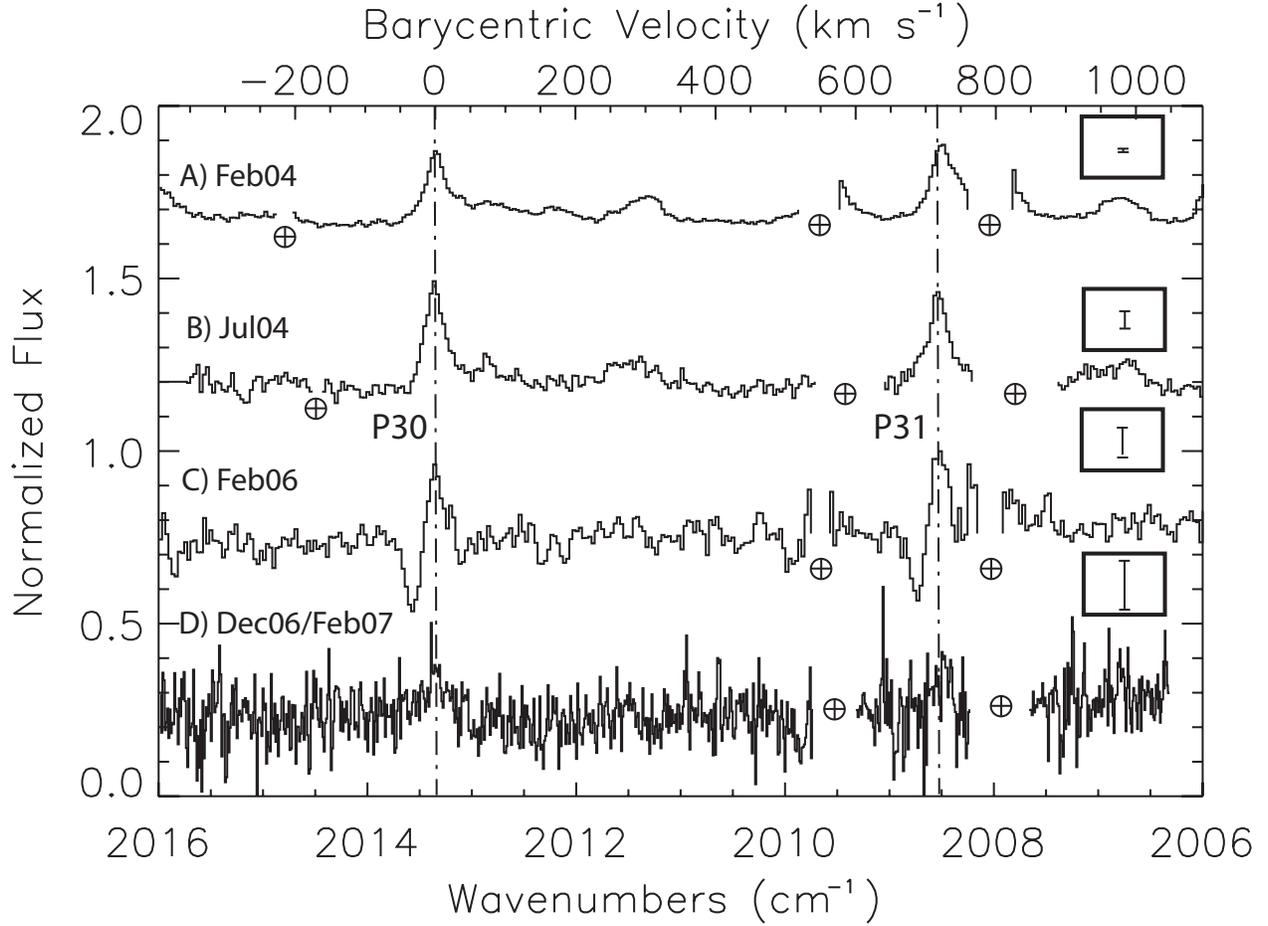}
\caption{Detailed comparison of the high-J CO lines. The v=1-0 P30 and P31 lines are plotted over four epochs spanning three years. An error bar has been plotted on the right side of each spectrum representing the standard deviation of the continuum. The spectra have been normalized, offset, and shifted to the rest frame of the molecular cloud in which V1647 Ori is embedded. }
\end{figure}
\clearpage

\begin{figure} 
\epsscale{0.8}
\plotone{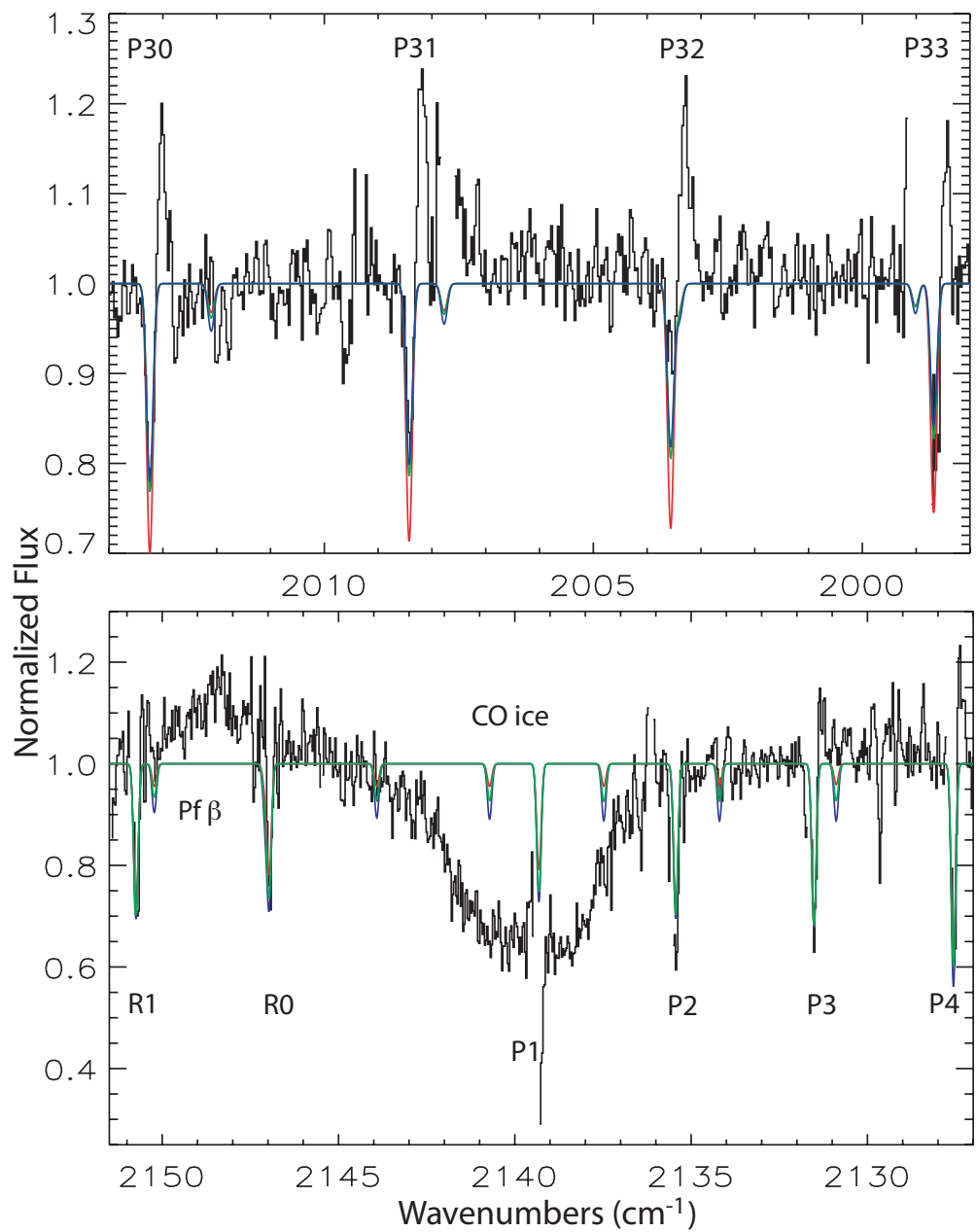}
\caption{Comparison of the v=1-0 CO spectrum after outburst (2006 February) and synthetic models. The appearance of blue-shifted absorption lines indicates the presence of an outflow.  We have plotted three synthetic spectra that bracket a reasonable range of parameters that describe this outflowing gas. The red, green and blue spectra correspond to gas temperatures of 1000K, 700K and 600K respectively.  The low temperature fits to the high-J lines are indistinguishable, however, the 1000K fit (red) is too deep. The CO ice feature at 2140 cm$^{-1}$ remains unchanged which is expected in light of its interstellar origin (Rettig et al. 2005) and is not modeled.} 
\end{figure}

\clearpage


\begin{thebibliography}


\bibitem[Andrews et al.(2004)]{2004ApJ...610L..45A} Andrews, S.~M., Rothberg, B., \& Simon, T.\ 2004, 
\apjl, 610, L45 
\bibitem[Aspin et al.(2006)]{2006AJ....132.1298A} Aspin, C., Barbieri, C., Boschi, F., Di Mille, F., Rampazzi, F., Reipurth, B., \& Tsvetkov, M.\ 2006, \aj, 132, 1298 
\bibitem[Balsara(2004)]{2004ApJS..151..149B} Balsara, D.~S.\ 2004, \apjs, 151, 149
\bibitem[Blake \& Boogert(2004)]{2004ApJ...606L..73B} Blake, G.~A., \& Boogert, A.~C.~A.\ 2004, \apjl, 606, L73 

\bibitem[Bohlin et al.(1978)]{1978ApJ...224..132B} Bohlin, R.~C., Savage,  B.~D., \& Drake, J.~F.\ 1978, \apj, 224, 132
\bibitem[Bouvier et al.(2007)]{2007prpl.conf..479B} Bouvier, J., Alencar, S.~H.~P., Harries, T.~J., Johns-Krull, C.~M., \& Romanova, M.~M.\ 2007, Protostars and Planets V, 479
\bibitem[Brice{\~n}o et al.(2004)]{2004ApJ...606L.123B} Brice{\~n}o, C., et al.\ 2004, \apjl, 606, L123 
\bibitem[Brittain et al.(2007)]{2007ApJ...659..685} Brittain, S. Simon, T., Najita, J. R., \& Rettig, T .W. \ 2007, \apj, 659, 685
\bibitem[Brittain et al.(2005)]{2005ApJ...626..283B} Brittain, S.~D., Rettig, T.~W., Simon, T., \& Kulesa, C.\ 2005, \apj, 626, 283
\bibitem[Brittain(2004)]{2004PhDT........21B} Brittain, S.~D.\ 2004, Ph.D.~Thesis,  
\bibitem[Brittain et al.(2003)]{2003ApJ...588..535B} Brittain, S.~D., Rettig, T.~W., Simon, T., Kulesa, C., DiSanti, M.~A., \& Dello Russo, N.\ 2003, \apj, 588, 535 
\bibitem[Brown et al.(2005)]{2005prpl.conf.8513B} Brown, J.~M., Boogert, A.~C.~A., Salyk, C., \& Blake, G.~A.\ 2005, Protostars and Planets V, 8513

\bibitem[Calvet et al.(1993)]{1993ApJ...402..623C} Calvet, N., Hartmann, L., \& Kenyon, S.~J.\ 1993, \apj, 402, 623 

\bibitem[Calvet et al.(1991)]{1991ApJ...383..752C} Calvet, N., Hartmann,  L., \& Kenyon, S.~J.\ 1991, \apj, 383, 752 

\bibitem[Camenzind(1990)]{1990RvMA....3..234C} Camenzind, M.\ 1990, Reviews in Modern Astronomy, 3, 234

\bibitem[Disanti et al.(2001)]{2001Icar..153..361D} Disanti, M.~A., Mumma, M.~J., Dello Russo, N., \& Magee-Sauer, K.\ 2001, Icarus, 153, 361 

\bibitem[Fedele et al.(2007)]{Fedetal}Fedele. D. van den Ancker, M.~E., Petr-Gotzens, M.~G. \& Rafanelli, P. \ 2007a, \aap, in press

\bibitem[Fedele et al.(2007)]{2007arXiv0707.0672F} Fedele, D., van den Ancker, M.~E., Petr-Gotzens, M.~G., Ageorges, N., \& Rafanelli, P.\ 2007b, ArXiv e-prints, 707, arXiv:0707.0672 

\bibitem[Gibb(1999)]{1999MNRAS.304....1G} Gibb, A.~G.\ 1999, \mnras, 304, 1

\bibitem[Gibb et al.(2006)]{2006ApJ...641..383G} Gibb, E.~L., Rettig, T.~W., Brittain, S.~D., Wasikowski, D., Simon, T., Vacca, W.~D., Cushing, M.~C., \& Kulesa, C.\ 2006, \apj, 641, 383 


\bibitem[Hartigan et al.(1995)]{1995ApJ...452..736H} Hartigan, P., Edwards, S., \& Ghandour, L.\ 1995, \apj, 452, 736

\bibitem[Hartmann et al.(2004)]{2004ApJ...609..906H} Hartmann, L., Hinkle, K., \& Calvet, N.\ 2004, \apj, 609, 906
\bibitem[Hartmann(1998)]{1998apsf.book.....H} Hartmann, L.\ 1998, Accretion processes in star formation, Cambridge University Press, 1998

\bibitem[Hartmann \& Kenyon(1996)]{1996ARA&A..34..207H} Hartmann, L., \& Kenyon, S.~J.\ 1996, \araa, 34, 207



\bibitem[Hinkle et al.(2003)]{2003SPIE.4834..353H} Hinkle, K.~H., et al.\ 2003, \procspie, 4834, 353 

\bibitem[Hinkle et al.(2000)]{2000SPIE.4008..720H} Hinkle, K.~H., Joyce, R.~R., Sharp, N., \& Valenti, J.~A.\ 2000, \procspie, 4008, 720 

\bibitem[Hinkle et al.(1998)]{1998SPIE.3354..810H} Hinkle, K.~H., Cuberly, R.~W., Gaughan, N.~A., Heynssens, J.~B., Joyce, R.~R., Ridgway, S.~T., Schmitt, P., \& Simmons, J.~E.\ 1998, \procspie, 3354, 810 

\bibitem[Johns-Krull \& Valenti(2000)]{2000ASPC..198..371J} Johns-Krull, C.~M., \& Valenti, J.~A.\ 2000, ASP Conf.~Ser.~198: Stellar Clusters and Associations: Convection, Rotation, and Dynamos, 198, 371

\bibitem[Kastner et al.(2006)]{2006ApJ...648L..43K} Kastner, J.~H., et al.\ 2006, \apjl, 648, L43 



\bibitem[Lada et al.(1991)]{1991ApJ...368..432L} Lada, E.~A., Bally, J., \& Stark, A.~A.\ 1991, \apj, 368, 432


\bibitem[McLean et al.(1998)]{1998SPIE.3354..566M} McLean, I.~S., et al.\ 1998, \procspie, 3354, 566 

\bibitem[McNeil et al.(2004)]{2004IAUC.8284....1M} McNeil, J.~W., Reipurth, B., \& Meech, K.\ 2004, \iaucirc, 8284, 1 

\bibitem[Mitchell et al.(2001)]{2001ApJ...556..215M} Mitchell, G.~F., Johnstone, D., Moriarty-Schieven, G., Fich, M., \& Tothill, N.~F.~H.\ 2001, \apj, 556, 215 


\bibitem[Muzerolle et al.(2005)]{2005ApJ...620L.107M} Muzerolle, J., Megeath, S.~T., Flaherty, K.~M., Gordon, K.~D., Rieke, G.~H., Young, E.~T., \& Lada, C.~J.\ 2005, \apjl, 620, L107 

\bibitem[Najita et al.(2000)]{2000prpl.conf..457N} Najita, J.~R., Edwards, S., Basri, G., \& Carr, J.\ 2000, Protostars and Planets IV, 457

\bibitem[Najita et al.(2003)]{2003ApJ...589..931N} Najita, J., Carr, J.~S., \& Mathieu, R.~D.\ 2003, \apj, 589, 931 


\bibitem[Ojha et al.(2006)]{2006MNRAS.368..825O} Ojha, D.~K., et al.\ 2006, \mnras, 368, 825 

\bibitem[Pontoppidan et al.(2002)]{2002A&A...393..585P} Pontoppidan, K.~M., Sch{\"o}ier, F.~L., van Dishoeck, E.~F., \& Dartois, E.\ 2002, \aap, 393,  585

\bibitem[Reipurth \& Aspin(2004)]{2004ApJ...606L.119R} Reipurth, B., \& Aspin, C.\ 2004, \apjl, 606, L119 

\bibitem[Rettig et al.(2006)]{2006ApJ...646..342R} Rettig, T., Brittain, S., Simon, T., Gibb, E., Balsara, D.~S., Tilley, D.~A., \& Kulesa, C.\ 2006, \apj, 646, 342 

\bibitem[Rettig et al.(2005)]{2005ApJ...626..245R} Rettig, T.~W., Brittain, S.~D., Gibb, E.~L., Simon, T., \& Kulesa, C.\ 2005, \apj, 626, 245 

\bibitem[Vacca et al.(2004)]{2004ApJ...609L..29V} Vacca, W.~D., Cushing, M.~C., \& Simon, T.\ 2004, \apjl, 609, L29 

\bibitem[Walter et al.(2004)]{2004AJ....128.1872W} Walter, F.~M., 
Stringfellow, G.~S., Sherry, W.~H., \& Field-Pollatou, A.\ 2004, \aj, 128, 
1872 

\end{thebibliography}
\end{document}